\documentclass[reprint, amsmath, amssymb, aps, prl,]{revtex4-1}
\pdfoutput=1

\usepackage[english]{babel}
\usepackage{silence}
\usepackage{graphicx}
\usepackage{braket}
\usepackage{dcolumn}
\usepackage{bm}
\usepackage{xcolor}
\usepackage[unicode=true,
 bookmarks=true,bookmarksnumbered=false,bookmarksopen=false,
 breaklinks=false,pdfborder={0 0 1},backref=false,colorlinks=true, linkcolor=magenta, urlcolor=blue, citecolor=blue, pdfstartview={FitH}, hyperfootnotes=true]
 {hyperref}
\makeatletter

\newcommand{\tr}{\text{Tr}}

\newcommand\scalemath[2]{\scalebox{#1}{\mbox{\ensuremath{\displaystyle #2}}}}

\newcommand{\im}{{\rm Im\,}}

\begin{document}

\preprint{APS/123-QED}

\title{Driven-dissipative Ising model: Dynamical crossover at weak dissipation}

\author{Daniel A. Paz}
\email[Corresponding author: ]{pazdanie@msu.edu}
\affiliation{Department of Physics and Astronomy, Michigan State University, East Lansing, Michigan 48824, USA}
\author{Mohammad F. Maghrebi}
\affiliation{Department of Physics and Astronomy, Michigan State University, East Lansing, Michigan 48824, USA}

\begin{abstract}
Driven quantum systems coupled to an environment typically exhibit effectively thermal behavior with relaxational dynamics near criticality. However, a different qualitative behavior might be expected in the weakly dissipative limit due to the competition between coherent dynamics and weak dissipation. In this work, we investigate 
a driven-dissipative infinite-range Ising model in the presence of individual atomic dissipation, a model that emerges from the paradigmatic open Dicke model in the large-detuning limit. We show that the system undergoes a dynamical crossover from relaxational dynamics, with a characteristic dynamical exponent $\zeta=1/2$, to underdamped critical dynamics governed by the exponent $\zeta=1/4$ in the weakly dissipative regime; a behavior that is markedly distinct from that of equilibrium. Finally, utilizing an exact diagrammatic representation, we demonstrate that the dynamical crossover to underdamped criticality is not an artifact of the mean-field nature of the model and persists even in the presence of short-range perturbations.
\end{abstract}

\maketitle

The search for new physics in non-equilibrium quantum systems has been fervently ongoing in recent years. A generic setting is provided by driven quantum systems coupled to the environment---also known as driven-dissipative systems. This setting has been realized in diverse platforms such as circuit QED \cite{fitzpatrick_observation_2017}, cavity QED \cite{baumann_dicke_2010, baden_realization_2014}, and trapped ions \cite{barreiro_open-system_2011, lin_dissipative_2013, safavi-naini_verification_2018}, among others, and has led to a flurry of theoretical efforts aiming to understand many-body open quantum systems far from equilibrium. These systems can harbor new, non-equilibrium phases of matter, and in the era of noisy intermediate-scale quantum devices \cite{preskill_quantum_2018}, offer an ideal setting to study how noise affects quantum systems. \par

It has become increasingly clear that generic driven-dissipative systems exhibit an effectively thermal behavior and relaxational dynamics near criticality
\cite{mitra_nonequilibrium_2006,wouters_absence_2006,diehl_quantum_2008,Gopalakrishnan10,nagy_critical_2011,Oztop12,DallaTorre12,torre_keldysh_2013, Sieberer13, marcuzzi_universal_2014,maghrebi_nonequilibrium_2016,foss-feig_emergent_2017,sieberer_keldysh_2016,Kilda19, kirton_introduction_2019}. 
This is because dissipation---together with the external drive---masks the underlying coherent dynamics. 
On the other hand, the dynamics could be qualitatively different in a weakly-dissipative system where incoherent effects occur infrequently \cite{lange_time-dependent_2018,lange_pumping_2017,rota_quantum_2019}; this is particularly relevant to systems where some (though small) degree of  dissipation is always present. 

In this work, we study the paradigmatic open Dicke model describing the collective interaction between atoms and a cavity mode, both subject to dissipation. This model hosts a phase transition to a superradiant phase with a macroscopic photonic population, which is observed in several experiments \cite{baumann_dicke_2010, baumann_exploring_2011, baden_realization_2014, klinder_dynamical_2015, zhiqiang_nonequilibrium_2017, safavi-naini_verification_2018, muniz_exploring_2020}. 
In the limit of large detuning, we show that the open Dicke model can be effectively described by the infinite-range driven-dissipative Ising model (DDIM) in a transverse field \cite{paz_long_paper}.
In contrast with the relaxational dynamics characteristic of driven-dissipative phase transitions, we show that this system undergoes a dynamical crossover to underdamped dynamics in the weakly-dissipative limit; the nature of this crossover is unique to driven-dissipative systems, and distinct from that of equilibrium in the absence of dissipation.
In addition to exact numerical simulation, we employ an analytical approach based on an extension of the Suzuki-Trotter quantum-to-classical mapping to the non-equilibrium realm of open quantum systems. The dynamical crossover might be viewed as an artifact of the mean-field character of the model; however, we show that it persists even in the presence of non-mean-field perturbations. Specifically, we consider perturbative short-range interactions which we can treat using a systematic diagrammatic representation.

\begin{figure}[t!]
    \centering
    \includegraphics[width=\linewidth]{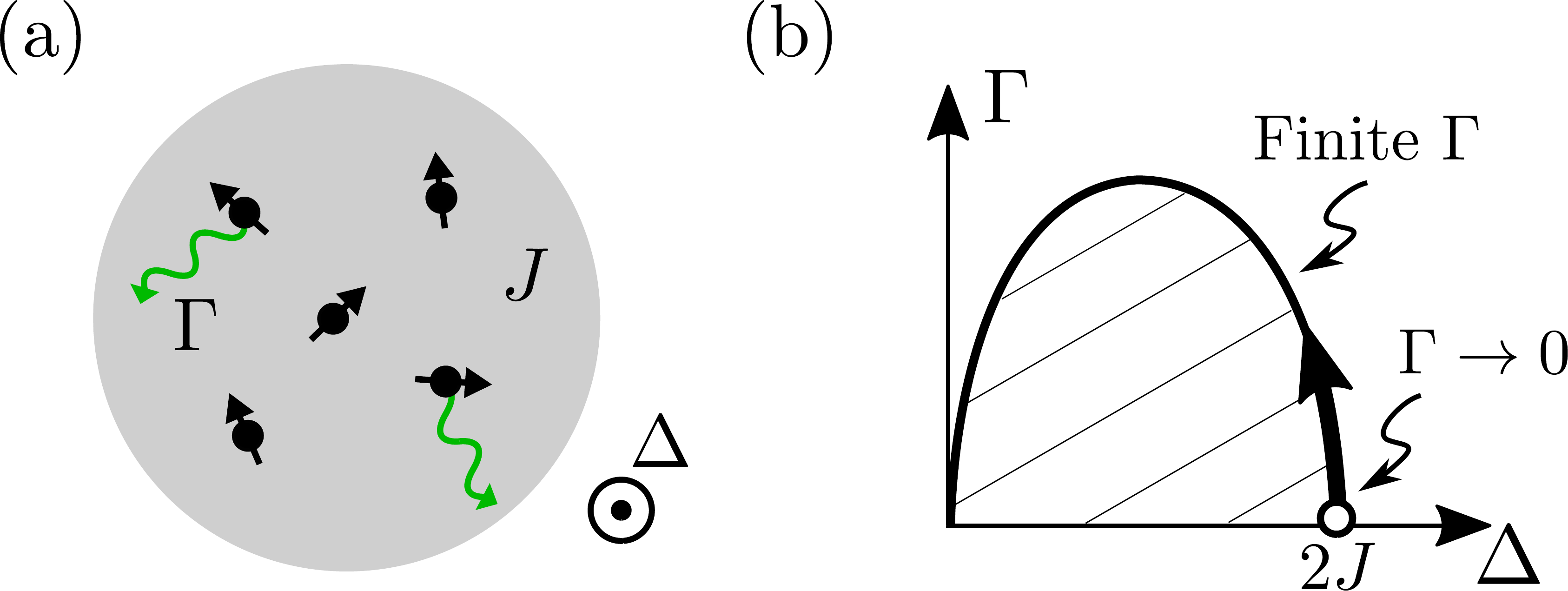}
    \caption{
    (a) Schematic depiction of the infinite-range driven-dissipative Ising model (DDIM) with $J$ the Ising interaction and $\Gamma$ the rate of atomic spontaneous emission. (b) Phase diagram of the DDIM. The shaded region denotes the ordered phase. The weakly dissipative critical point of the DDIM, $\Gamma \to 0$, exhibits underdamped dynamics in contrast with the relaxational dynamics at a generic critical point. The weakly dissipative critical point is an unstable fixed point with respect to dissipation, giving rise to a dynamical crossover.
    }
    \label{phasespace}
\end{figure}

\textit{Model}.---The open Dicke model has become one of the quintessential models of open quantum systems \cite{kirton_introduction_2019}. It is characterized by a system of 2-level atoms that are driven externally, experience spontaneous emission, and are coupled to a single-mode lossy cavity. In the rotating frame of the drive, the coherent dynamics is governed by the Hamiltonian 
\begin{equation}\label{ODM Hamiltonian}
    H_{\text{Dicke}} = \omega_0 a^\dagger a + \Delta S_z + \frac{2g}{\sqrt{N}}S_x(a + a^\dagger)\,,
\end{equation}
with $g$ the atom-cavity coupling strength, $\omega_0$ the cavity detuning, and $\Delta$ the atomic level-splitting. We have also defined the total spin operators $S_\alpha = \sum_i \sigma^\alpha_i$ where $\sigma^\alpha$ for $\alpha \in \{x,y, z\}$ are the usual Pauli matrices. The incoherent dynamics is described by photon loss at a rate $\kappa$ together with spontaneous emission of individual spins at a rate $\Gamma$. In the limit of large detuning and cavity loss, the cavity mode can be adiabatically eliminated \cite{morrison_dynamical_2008, damanet_atom-only_2019, keeling_collective_2010, paz_long_paper, luo_dynamic_2016}, and the model can be described by an effective driven-dissipative infinite-range Ising model in a transverse field [schematically shown in Fig.~\ref{phasespace}(a)],
\begin{align}\label{Hamiltonian}
 \begin{split}
 \frac{d \rho}{d t}=   &\mathcal{L}(\rho) \equiv
  -i [H, \rho] +  \Gamma\sum_i \mathcal{D}_{\sigma^-_i}[\rho], \\
&\mbox{with} \quad  H = -\frac{J}{N}S_{x}^2 + \Delta S_{z},
\end{split}
\end{align}
where $\mathcal{D}_L(\rho)=L \rho L^\dagger - \frac{1}{2}\{L^\dagger L, \rho\}$ represents dissipation;
see \cite{paz_long_paper} for the microscopic values of $J, \Delta$.
While we focus on the effective spin model, we emphasize that our results are applicable to the open Dicke model. Just like the Dicke model, the DDIM has a $\mathbb{Z}_2$ symmetry under $\sigma^{x,y} \to -\sigma^{x,y}$ (which does not imply a conserved quantity for open systems \cite{albert_symmetries_2014}). This symmetry is spontaneously broken in the transition from the normal ($\langle S_x\rangle = 0$) to the ordered ($\langle S_x \rangle \neq 0$) phase. \par 

We represent the spin model in an exact field-theoretical description amenable to analytical treatment. This is achieved by exactly mapping the non-equilibrium partition function
\begin{equation*}
    Z = \tr[\rho(t)] = \tr\left[e^{\mathcal{L}t} (\rho_0)\right],
\end{equation*}
to a scalar field theory. (While $Z=\tr(\rho)=1$, quantities of interest can be computed by inserting sources \cite{sieberer_keldysh_2016}.) To this end, we adapt the Suzuki-Trotter decomposition \cite{suzuki_relationship_1976} to this non-equilibrium setting \cite{paz_long_paper}. The Ising interaction in the resulting classical action is then conveniently decoupled via a Hubbard-Stratonovich transformation. The result is a path integral representation of the non-equilibrium partition function \cite{paz_long_paper}
\begin{equation}
    Z = \int \mathcal{D}[m^{(u/l)}]e^{i\mathcal{S}[m^{(u)}(t),m^{(l)}(t)]}\,,
\end{equation}
where $m^{(u/l)}$ are the Hubbard-Stratonovich fields, $u/l$ the ``forward/backward'' branches of the Keldysh contour \cite{kamenev_field_2011}, and $\mathcal{D}[m^{(u/l)}]$ is the path-integral measure containing unimportant prefactors left over from the quantum-to-classical mapping. The action, which is exact, is given by
\begin{equation}\label{DDIMaction}
\begin{split}
  \null\!\!\!\!\mathcal{S} =  -2JN\!\int_t m_{c}(t)m_{q}(t) 
  -iN\ln\tr\Big[\mathcal{T}e^{\int_t \mathbb{T}(m_{c/q}(t))}\Big]\,.
\end{split}
\end{equation}
Here, $\mathcal{T}$ denotes time ordering, and the ``classical/quantum'' fields  $m_{c/q} = (m^{(u)} \pm m^{(l)})/\sqrt{2}$ are defined for future convenience, and the matrix $\mathbb{T}$ is given by
\begin{equation}
    \mathbb{T} = 
    \renewcommand*{\arraystretch}{1.5}
    \scalemath{.69}{
    \begin{pmatrix}
    -\frac{\Gamma}{4}+i2\sqrt{2}J m_q & i\Delta & -i\Delta & \frac{\Gamma}{4} \\
    i\Delta - \frac{\Gamma}{2} & -\frac{3\Gamma}{4} + i 2\sqrt{2} J m_c  & -\frac{\Gamma}{4} & -i\Delta -\frac{\Gamma}{2} \\
    -i\Delta - \frac{\Gamma}{2} &-\frac{\Gamma}{4} &- \frac{3\Gamma}{4} -i2\sqrt{2} J m_c & i\Delta - \frac{\Gamma}{2} \\
    \frac{\Gamma}{4} & -i\Delta & i\Delta & -\frac{\Gamma}{4} -i 2\sqrt{2} J m_q
    \end{pmatrix}
    },
\end{equation}
with the time-dependence of the fields $m_{c/q}$ made implicit.  
This formalism in terms of a single scalar field (reflecting the Ising symmetry) provides an intriguing description particularly in the presence of individual atomic dissipation where existing (Holstein-Primakoff \cite{torre_keldysh_2013} or fermionic \cite{dalla_torre_dicke_2016}) techniques are either inapplicable or rather complex. Finally, the collective nature of the Ising interaction leads to an overall factor of $N$ in the action, enabling us to obtain exact results from the saddle-point approximation.

\textit{Critical Properties \& Finite-Size Scaling}.---We first compute the magnetization through the saddle-point approximation, exact in the thermodynamic limit $N\to \infty$. To this end, we set
\(    
\delta \mathcal{S}/{\delta m_{c/q}(t)} = 0
\) 
and seek a solution with  $m_q(t) = 0$ and $m_c(t) \equiv m=$ const. Besides the normal phase (for $\Gamma > \Gamma_c\equiv 4\sqrt{\Delta(2J-\Delta)}$) with the trivial solution $m = 0$, an ordered phase emerges (when $\Gamma < \Gamma_c$) with two nontrivial stable solutions, $m =\pm \sqrt{-\Gamma^2-16\Delta^2+32\Delta J}/4J$, signifying the breaking of the $\mathbb{Z}_2$ symmetry; see the phase diagram in Fig. \ref{phasespace}(b).
Notice the absence of ordering at $\Delta=0$ as can be generally proved in driven-dissipative  Ising-type models \cite{foss-feig_solvable_2017}.
To characterize fluctuations, we expand the action in powers of fluctuations around $m_{c/q} = 0$. Within the normal phase. We have, up to the quadratic order,
\begin{equation}
\label{action expansion}
      \mathcal{S}^{(2)} = \frac{1}{2}\int_{t, t'} \begin{pmatrix}m_c,\,
      m_q\end{pmatrix}_{t}
      \begin{pmatrix}0 & P^A\\ P^R & P^K\end{pmatrix}_{t-t'}
      \begin{pmatrix}m_c\\m_q\end{pmatrix}_{t'},
\end{equation}
with 
\begin{align}\label{inverse functions}
P^{R}(t)&=P^{A}(-t)= -2J\delta(t) + \Theta(t)8J^2 e^{-\frac{\Gamma}{2} |t|}\sin{(2\Delta t)}, \nonumber \\
    P^K(t) &= i8J^2e^{-\frac{\Gamma}{2} |t|}\cos{(2\Delta t)}.
\end{align}
In a slight abuse of notation, a factor of $\sqrt{N}$ has been absorbed into both $m_c$ and $m_q$. 
The above action is \textit{exact} in the large-$N$ limit away from the critical point---higher order terms are suppressed as ${\cal O}(1/N)$. We can then investigate the auto-correlation and spectral response functions:
\begin{align*}\label{correlation function}
C(t) &= \frac{1}{N}\langle\{S_x(t), S_x(0)\}\rangle =  \langle m_c(t) m_c(0) \rangle\,, \\
      \chi(t) &= \frac{1}{i N}\langle[S_x(t), S_x(0)]\rangle
    = \frac{1}{i}\langle m_q(t) m_c(0)
    - m_c(t) m_q(0) \rangle.
\end{align*}
These functions can be obtained by inverting the kernel in Eq. \eqref{action expansion}. For the exact relationship between the fields $m_{c/q}$ and the spin operator $S^x$, see Ref.~\cite{paz_long_paper}.\par

To identify the critical properties, we define the distance from the critical point $\Gamma_c$ as $\gamma = \Gamma - \Gamma_c$.
A simple analysis close to the critical point ($\gamma \ll \Gamma$) at long times ($t \gg 1/\Gamma$) yields the correlation and response functions 
\begin{equation}\label{GPscaling}
C(t) \sim \frac{16 J\Delta}{\gamma\Gamma_c}e^{-\gamma |t|/2}, \quad \chi(t)  \sim {\rm sgn}(t)\frac{- 4\Delta}{\Gamma_c}e^{-\gamma |t|/2}.
\end{equation}
The critical behavior near the phase transition can be characterized by an effective temperature via the \textit{classical} fluctuation-dissipation relation, $\chi(t) = \partial_t C(t)/2T_{\rm eff}$ \cite{tauber_critical_2014}. 
Equation~\eqref{GPscaling} then reveals that $T_{\rm eff} = J$ everywhere along the phase boundary.
Note that correlations diverge as $1/\gamma$ upon approaching the critical point, $\gamma\to0$, while the response function remains finite. 
These scaling properties determine the scaling dimensions of the fields, upon rescaling time (or inversely rescaling $\gamma$), as $[m_c] = \frac{1}{2}$ and $[m_q] =  -\frac{1}{2}$. These are consistent with the relaxational dynamics of the mean-field classical Ising model \cite{hohenberg_theory_1977}.\par 

Next, we turn our attention to the weakly dissipative critical point where $\Gamma \to 0$ and $\Delta=2J$; see Fig.~\ref{phasespace}(a).
In this limit, dissipation is only acting infrequently, and distinct qualitative behavior may be expected. 
We find (for arbitrary $\Gamma t$)
\begin{align}
\begin{split}
    &C(t) \sim \frac{32J^2}{\Gamma^2}(2 + \Gamma |t|)e^{-\Gamma |t| /2},\\
    &\chi(t) \sim -8 J t\,e^{-\Gamma |t|/2}.
\end{split}
\end{align}
Correlations now diverge as $1/\Gamma^2$ upon approaching the weakly dissipative critical point $\Gamma\to0$ for fixed $t$. In addition, the response function scales linearly with $t$. These observations yield the scaling dimensions $[m_c] = 1$ and $[m_q] = 0$,
which are distinct from those at a generic critical point discussed above. Indeed, one can see that $\im P^R(\omega) \sim \Gamma \omega$ at low frequencies, therefore the effective dissipation vanishes in the limit $\Gamma \to0$, a fact that underscores its distinct scaling behavior.
We shall see shortly that the weakly dissipative critical point is described by underdamped, rather than overdamped, dynamics. 

\begin{figure}[t!]
    \centering
    \includegraphics[width=\linewidth]{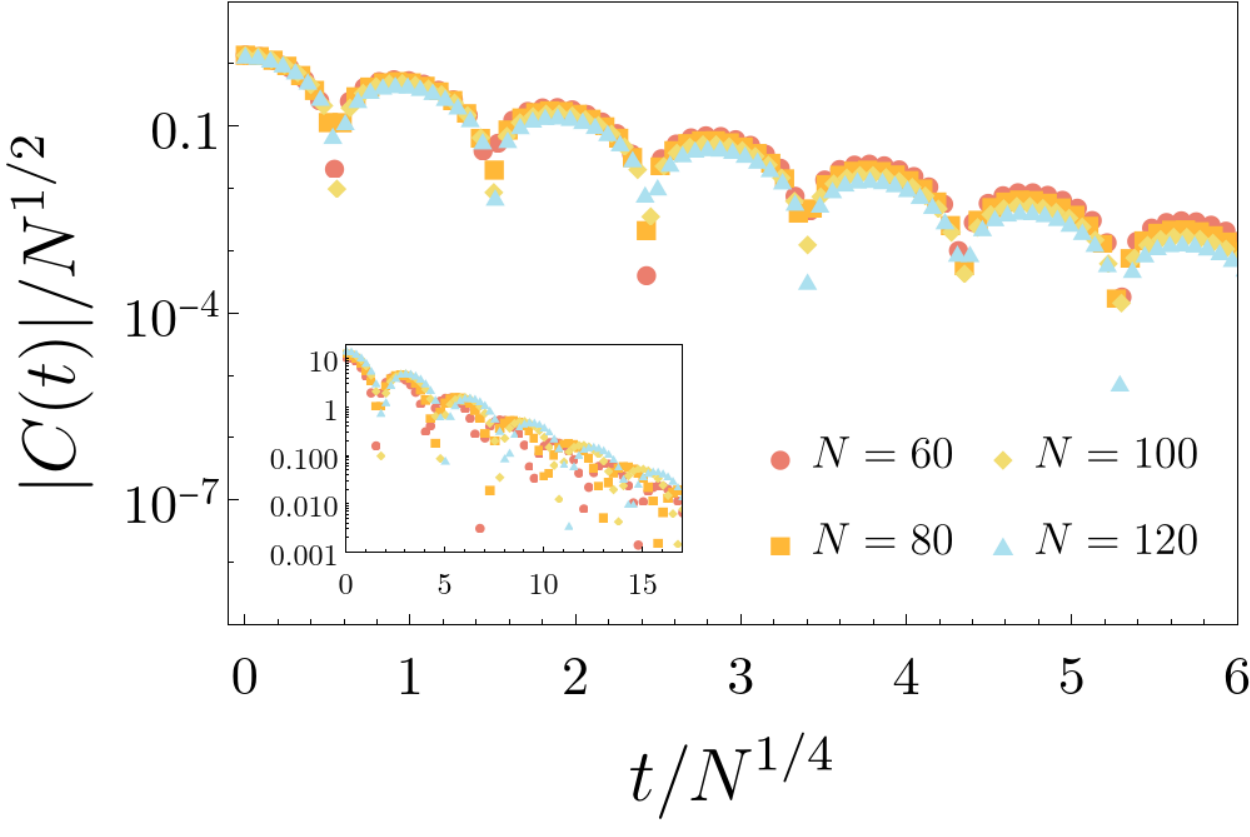}
    \caption{Finite-size scaling of the correlation function near the weakly dissipative critical point, where $\Delta$ is tuned such that we lie along the phase boundary ($J = 1, \Gamma = 0.5$). 
    The dynamics is underdamped in contrast with the purely relaxational behavior at a generic driven-dissipative phase transition, and exhibits the critical scaling $t \sim N^{1/4}$ to be contrasted with $t \sim N^{1/2}$ of relaxational dynamics. The inset shows the unscaled plots for comparison. Underdamped universal dynamics persist for $t \gg 1/\Gamma$.}
    \label{SP FS scaling plot}
\end{figure}

So far, we have inspected the critical behavior near, but away from, the critical point. The divergence of the correlation function at the critical point will be regularized by the finite size of the system, but requires finite-size scaling. 
To this end, we first remark that the most relevant interaction term in the action is given by the ``classical vertex''  $\sim\frac{1}{N}\int_t m_c^3(t) m_q(t)$ as $[m_c]>[m_q]$. Upon rescaling time ($t \to \lambda t$), this term remains invariant only if $N$ is rescaled appropriately. Using the scaling dimensions of the fields, we find that $N$ has to be rescaled as $N\to \lambda^2 N$ at a generic critical point, while $N\to \lambda^4 N$ at the weakly dissipative critical point. The correlation function at criticality can then be written as 
\begin{equation}\label{fs scaling}
    C(t) =N^\alpha {\cal C}(t/N^\zeta)\,,
\end{equation}
where the exponent $\alpha$ dictates the scaling of fluctuations, and the exponent $\zeta$ defines a dynamical critical exponent. While \textit{static} fluctuations always scale as $N^{1/2}$, identifying $\alpha=1/2$, the generic and weakly dissipative critical points are characterized by two distinct \textit{dynamical} exponents, $\zeta=1/2$ and $\zeta=1/4$, respectively. Furthermore, in contrast with the purely relaxational dynamics at a generic critical point \cite{paz_long_paper}, the weakly dissipative regime exhibits underdamped critical dynamics (Fig.~\ref{SP FS scaling plot}).  \par
\begin{figure}[t!]
    \centering
    \includegraphics[scale=.5]{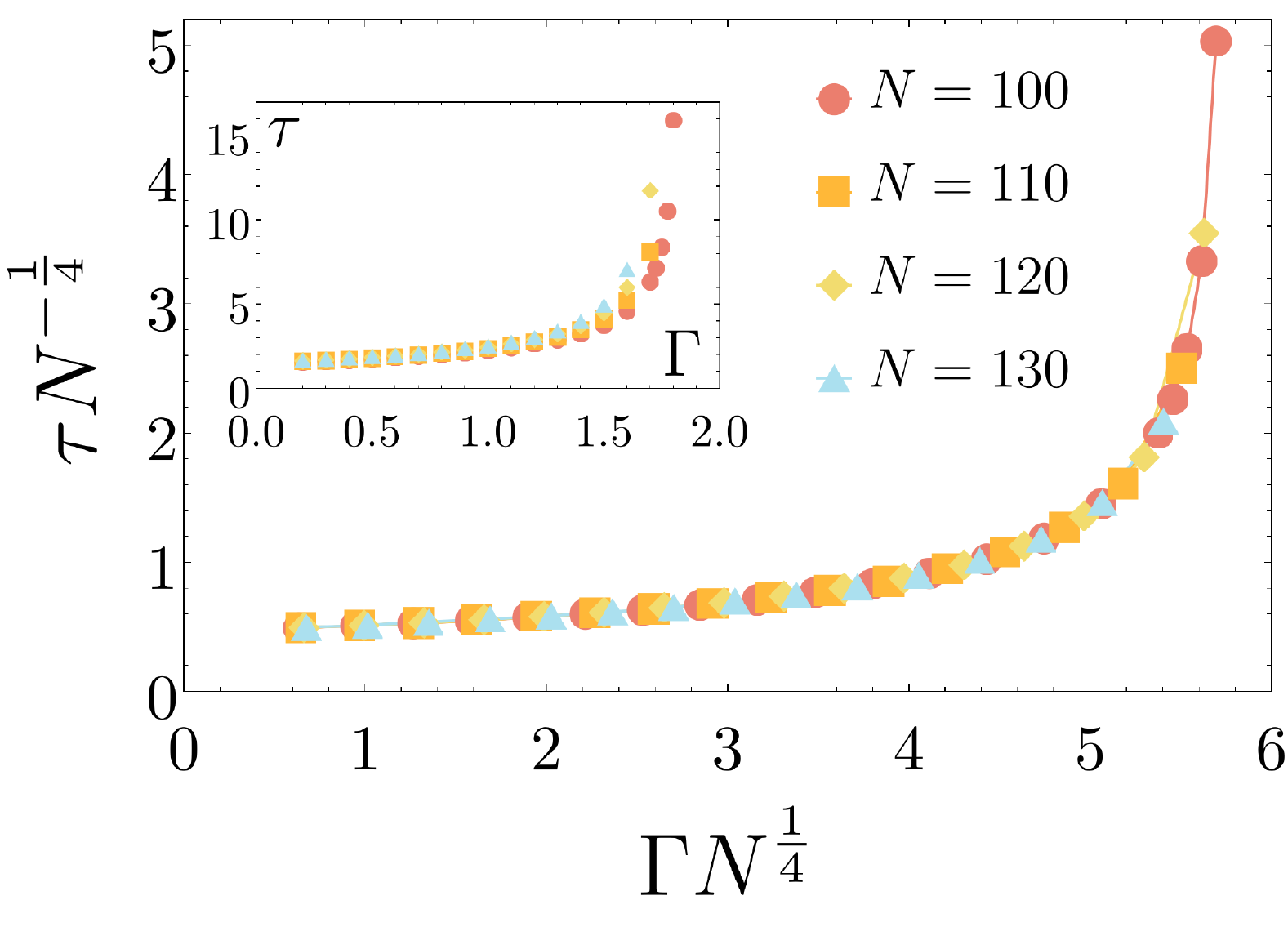}
    \caption{First zero of correlation function $\tau$ as a function of $\Gamma$ and for different system sizes.
    For any value of $\Gamma$, we choose $\Delta$ such that we remain on the phase boundary. Above a threshold of $\Gamma N^{1/4} \approx 6$, the time $\tau$ diverges, signaling the onset of the crossover from underdamped to overdamped dynamics. The inset shows the unscaled plots for comparison.}
    \label{crossover}
\end{figure}

 
The underdamped criticality in the weakly dissipative regime does not only emerge in the extreme limit $\Gamma \to 0$, but rather more generally for moderate values of $\Gamma$ depending on system size.
In Fig.~\ref{SP FS scaling plot}, we show the underdamped dynamics and a universal dynamical scaling for $\Gamma/J = 0.5$ where $J t\lesssim 20$ and $N\lesssim 120$. Remarkably, this critical behavior persists, for the same system sizes, up to $\Gamma \gtrsim 1$ (while keeping on the phase boundary). The underdamped critical dynamics persists even at long times compared to 1/$\Gamma$ as one can see from Fig.~\ref{SP FS scaling plot}, therefore it should not be viewed as an artifact of short times where dissipation has not acted.
As $\Gamma$ is further increased, the relaxational dynamics eventually sets in, and thus the system exhibits a dynamical crossover between distinct dynamical criticalities. We can identify this crossover as a function of $\Gamma$ and the system size $N$. Indeed, a simple scaling analysis (upon rescaling $t\to \lambda t$) requires $\Gamma \to \Gamma/\lambda$ at the weakly dissipative critical point, resulting in a flow equation $d\Gamma/dl=\Gamma$ where we substituted $\lambda=e^{-l}$. This identifies $\Gamma_*=0$ as an unstable fixed point, as schematically shown in Fig. \ref{phasespace}(b). Together with finite-size scaling, $N \to \lambda^4N$, at the weakly dissipative critical point, we are led to a general scaling form for the time scale associated with underdamped oscillations as 
\begin{equation}
    \tau = N^{1/4} \hat \tau(\Gamma N^{1/4}),
\end{equation}
where $\hat \tau$ is a general scaling function. 
We confirm this scaling behavior in Fig. \ref{crossover} where we have identified $\tau$ as the first zero of the correlation function. Notice that this time scale diverges around $\Gamma N^{1/4}\approx 6$, signifying the onset of the dynamical crossover.
A similar crossover emerges in equilibrium as the temperature is turned on; however, the interplay between unitary and dissipative dynamics sets the nature of this crossover and even the critical exponents apart from those in equilibrium \cite{paz_long_paper}.

\textit{Beyond mean-field models}.---The open Dicke model and its spin-only Ising variant are special due to their mean-field character.
An immediate question is to what extent the dynamical features discussed here persist in the presence of generic (non-mean-field) perturbations such as short-range interactions. Such perturbations couple the collective field $m$ to ``spin waves'' $ \widetilde{\sigma}^x_k = \sum_j e^{-ikj}\sigma_j^x$ for $k\ne 0$. One can show that these spin waves are effectively at the same temperature as the order parameter \cite{paz_long_paper}. Therefore, $m$ can be viewed as a ``macroscopic'' degree of freedom coupled to a large thermal bath of spin waves (with $\sim N$ modes). The conventional wisdom---borrowed from the Caldeira-Leggett model \cite{caldeira_quantum_1983}, for example---is that the thermal bath induces dissipation on the dynamics of $m$. Of course, the Markovian bath already leads to dissipative dynamics at any finite $\Gamma$ even in the absence of spin waves; however, in the limit $\Gamma \to 0$, spin-wave-induced dissipation could drastically change the nature of the dynamics and spoil the underdamped critical behavior. Surprisingly, this is not the case and the dynamical crossover is robust against non-mean-field perturbations, as we show next. To be concrete, let us consider a perturbative nearest-neighbor interaction,
\begin{equation}
    H_{\text{NN}} = H - \lambda \sum_i \sigma_i^x \sigma^x_{i+1}\,,
\end{equation}
assuming $\lambda \ll J$. Using our quantum-to-classical mapping, we can systematically treat spin waves in a field-theoretical framework where $m_k$ represents spin waves that emerge from the Hubbard-Stratonovich transformation of the short-range Hamiltonian \cite{paz_long_paper}. To find the spin-wave-induced dissipation, we need to characterize the self energy $\Sigma^R(\omega)$; this quantity provides the first correction to $P^R(\omega)$ due to the nonlinear coupling to spin waves.
\begin{figure}
    \centering
    \includegraphics[width=\linewidth]{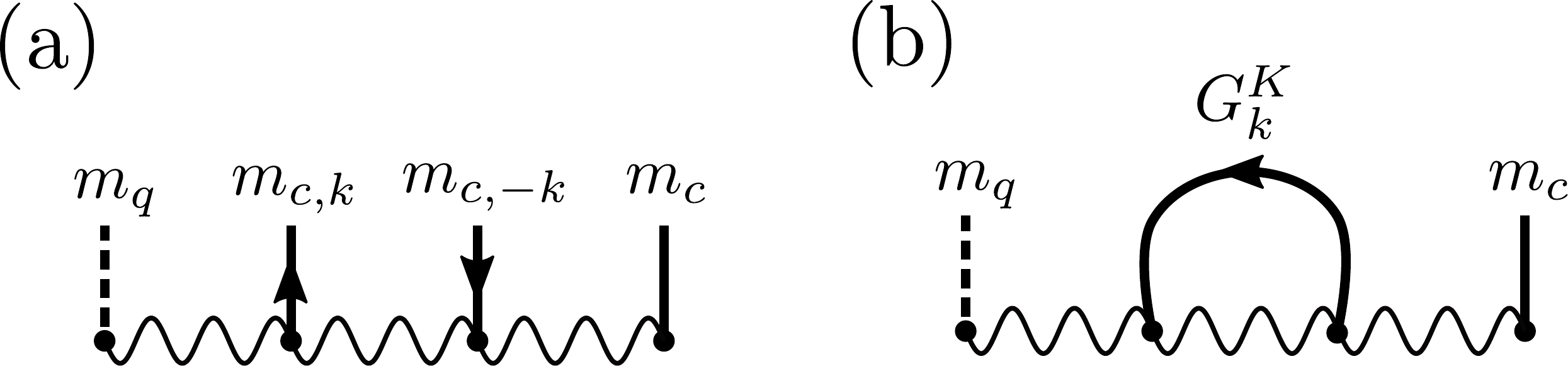}
    \caption{(a) A representative diagram showing a quartic interaction vertex between the collective mode $m_{c/q}$ and the spin waves $m_{c,k}$ with momentum $k\ne 0$. This diagram is time ordered from right to left. (b) The one-loop contribution of the diagram in (a) to the self energy $\Sigma^R$; here, $G^K_k$ is the Keldysh Green's function corresponding to spin waves \cite{paz_long_paper}.}
    \label{diagrams}
\end{figure}%
We compute the self energy perturbatively in powers of $\lambda$ using a diagrammatic representation \cite{paz_long_paper}. A representative diagram, and its contribution to the self energy, is shown in Fig. \ref{diagrams}. We report the details of this analysis in Ref.~\cite{paz_long_paper}, and just quote the result at one loop (to the order ${\cal O}(\lambda^2$)) in the low-frequency limit:
\begin{equation}
\begin{split}
    \Sigma^R(\omega)&=  \left[\frac{1536 J^2 \Delta }{(\Gamma^2 + 16\Delta^2)^2} + \frac{ 8192 J^2 \Delta \Gamma}{(\Gamma^2 + 16\Delta^2)^3}i \omega \right] \lambda^2 \,. 
\end{split}
\end{equation}
Interestingly, $\im \Sigma^R(\omega)$ vanishes as $\Gamma \to 0$, which  means that the coupling to a thermal bath of spin waves does not induce dissipation in this limit. We thus conclude that the underdamped critical behavior is robust against short-range perturbations, at least to the order $\lambda^2$, and is not just a mean-field artifact. 
Finally, we remark that the real part of $\Sigma^R$ renormalizes the critical point, which in turn makes the ordered region of the phase diagram shrink, as should be expected \cite{zhu_dicke_2019, lerose_chaotic_2018,paz_long_paper}.

\textit{Summary and Outlook}.---In this work, we have considered an experimentally relevant limit of the open Dicke model, namely the driven-dissipative Ising model with infinite-range interactions. We have shown that the critical dynamics becomes underdamped in the weakly dissipative regime, in contrast with the relaxational dynamics at a generic critical point. We have further identified the dynamical crossover and its distinct nature from that of equilibrium.
Finally, we have shown that the underdamped critical behavior persists in the presence of short-range perturbrations, and thus is not an artifact of the mean-field nature of the model considered here.
Extension of this work to the full open Dicke model outside the domain of adiabatic elimination is a natural direction for future research. The diagrammatic techniques used to treat short-range perturbations should prove valuable in studying the short-range version of Eq. \eqref{Hamiltonian}, and will be investigated in future work.

\textit{Acknowledgements}.---This material is based upon work supported by the NSF under Grant No. DMR-1912799. The authors also acknowledge the start-up funding from Michigan State University. We are also indebted to Alireza Seif and Paraj Titum for valuable discussions.

\end{document}